\documentstyle[aps,draft,epsf]{revtex} 

\def\lesim{\,{\raise-3pt\hbox{$\sim$}}\!\!\!\!\!{\raise2pt\hbox{$<$}}\,}

\begin{document}

%
%

\title{ Sources of UHE Neutrinos}
 \vspace{1.3cm}
\author{Mou Roy and H.J. Crawford\\ }
\address{ MS 50-245, Lawrence Berkeley National Laboratory,\\
         1 Cyclotron Road, Berkeley, CA 94720.}
\preprint{}
\date{\today}
\baselineskip 21 pt

\maketitle

\begin{abstract}

\noindent
In this paper we give a brief systematic study
of different models of astrophysical point sources of 
ultra high energy ($ \ge 1 $ TeV) neutrinos.

\end{abstract}

\newpage

\bigskip\bigskip

\section{Introduction}

We look at possible ultra high energy
(UHE) neutrino point sources, galactic and extragalactic,
focusing in particular on 
neutrinos of energies $\ge $ 1 TeV.
We present neutrino fluxes from all sources 
in common units.
Owing to their small interaction cross section, 
neutrinos arrive undeflected from the source,
carrying information about extremely high energy processes at
cosmological distances.
Recent detection of UHE photons from different astrophysical
sources indicate that UHE neutrinos could also be
produced provided the photons 
are of hadronic origin,
via the $ \pi ^\pm \rightarrow \mu^\pm
\rightarrow e^\pm$ decay chain. 

A brief outline of the paper is as follows:
Section II illustrates
Active Galactic Nuclei  models for high energy
neutrino emission. Section III describes feasible scenarios
where such high energy neutrinos could emanate from supernova
remnants and pulsars, also emphasizing  possible nuclear
interactions.
In Section IV we look at TeV neutrino emission from two
recent $\gamma$ ray burst models followed by the discussion of topological
defect models predicting UHE neutrino emission in Section V.
In Section VI we give the conclusions and an overview.

\section{Ultra high energy neutrinos emanating from AGN}

Active Galactic Nuclei
(AGN) are the most luminous  objects in the Universe, their
luminosities ranging from $10^{42}$ to $10^{48}$ ergs$/$sec.
They are believed to be powered by  central supermassive black holes
whose masses are of the order of $10^4 $ to $10^{10} M_{\odot}$.
The detection of ultrahigh energy neutrinos from AGNs would
confirm the hadronic model associated with high energy
$\gamma $ ray production since there is no source for high
energy neutrinos in electromagnetic models.
The production of such neutrinos can be described in terms of the spherical 
accretion AGN model \cite{sph} or the blazar model 
\cite{mannheimb,protherb}.
However, since the observed UHE gamma rays ($\ge$ 1 TeV)
by the Whipple, HEGRA and
CANGAROO observatories originate from known blazars 
(Mkn 421 and Mkn 501), the favored model for UHE neutrino
production would be the blazar model of AGN \footnote{Also the 
second EGRET catalog of high-energy gamma ray sources \cite{thom}
contains 40 high confidence identifications of AGN, and all appear to
be blazars.}. Most recent data indicates $\gamma$ ray energy spectra up to
5-8 TeV from Mkn 421 and $\sim 10$ TeV from Mkn 501 without any well
defined cutoff \cite{recent}.
However, spherical accretion
instrumental in the production of UHE particles should not
be ruled out completely. Recent models predict our
galaxy to harbor a $\sim 10^{6} M_{\odot}$ black hole
at the center which could emit such neutrinos; a theoretical explanation
of this scenario would be similar to the spherical accretion
model of AGN as we have discussed below.

\subsection{Blazar Models for Ultrahigh Energy Neutrino Production}

Models predicting $\gamma$ ray emission from AGN jets are primarily leptonic
in nature.
Most of these models assume that the $\gamma$ rays are
produced by inverse Compton scattering off electrons contained in a blob of
plasma which itself is in relativistic motion in the general direction
of the observer \cite{agnlep}. The most relevant process is assumed to be
the SSC (Synchrotron Self Compton) process in which high energy gamma rays 
are produced by
relativistic electrons via Compton scattering off synchrotron photons.
If this is the mechanism, the radiation energy density must dominate
the magnetic energy density. 
Such leptonic models do not predict the production of UHE neutrinos.
However, there are hadronic models for
particle emission from blazars which are consistent with all the data.
Such models have been
proposed by Mannheim \cite{mannheimb} and Protheroe \cite{protherb}
independently. 
Photoproduction by shock accelerated protons in a relativistic
jet at some distance from the central black hole is assumed by 
both models, though these two models assume different target
spectra. Mannheim assumes a sychrotron spectrum whereas Protheroe
assumes a disk spectrum.
A brief description
and some saliant features of each are given below.

\subsubsection{Proton Initiated Cascade (PIC) model}

In the PIC (Proton Initiated Cascade) model \cite{mannheimb} relativistic 
shocks are 
hypothesised to be propagating
down an expanding jet. 
Particle acceleration at relativistic and oblique 
shocks by diffusive and
drift mechanisms has been studied in detail in the past \cite{drift}.
Protons obey such mechanisms and upon entering the jet
are accelerated to energies 
where they initiate cascades due to photoproduction ($ p + \gamma 
\rightarrow \pi^0, \pi^\pm...$).
The crucial idea is that shock acceleration theory gives us a
relativistic proton population and also provides soft synchrotron
photons from accelerated electrons which act as  `targets' for production
of pions and pairs. Ultrahigh energy neutrinos ($\ge$ TeV) and 
photons are produced from
pion decays.
Another crucial requirement for these models is that the 
acceleration is faster than
expansion and the Larmor radius of gyrating particles does not exceed
the radius of curvature of the shocks \cite{bell}.
Considering first order Fermi acceleration across the shock the maximum
achievable proton energy is calculated to be $\sim 10^6 $ TeV for
a typical AGN jet magnetic field value of $1$ G \cite{mannheimb}. 
The corresponding maximum Larmor radius would be $\sim 10^{15}$ cm.
If observed hot spots in active galaxies
 are believed to be termination shocks of
jets carrying energy from active nuclei to the lobes, a typical value
of the maximum radius of the jet would be $\sim 10^{21}$ cm, which is the
maximum radius of curvature of the shock \cite{bell}. The neutrino
production site would then be well within the jet as described in 
both the blazar models investigated here.

\setbox4=\vbox to 160 pt {\epsfysize= 5 truein\epsfbox[0 -200 612 592]{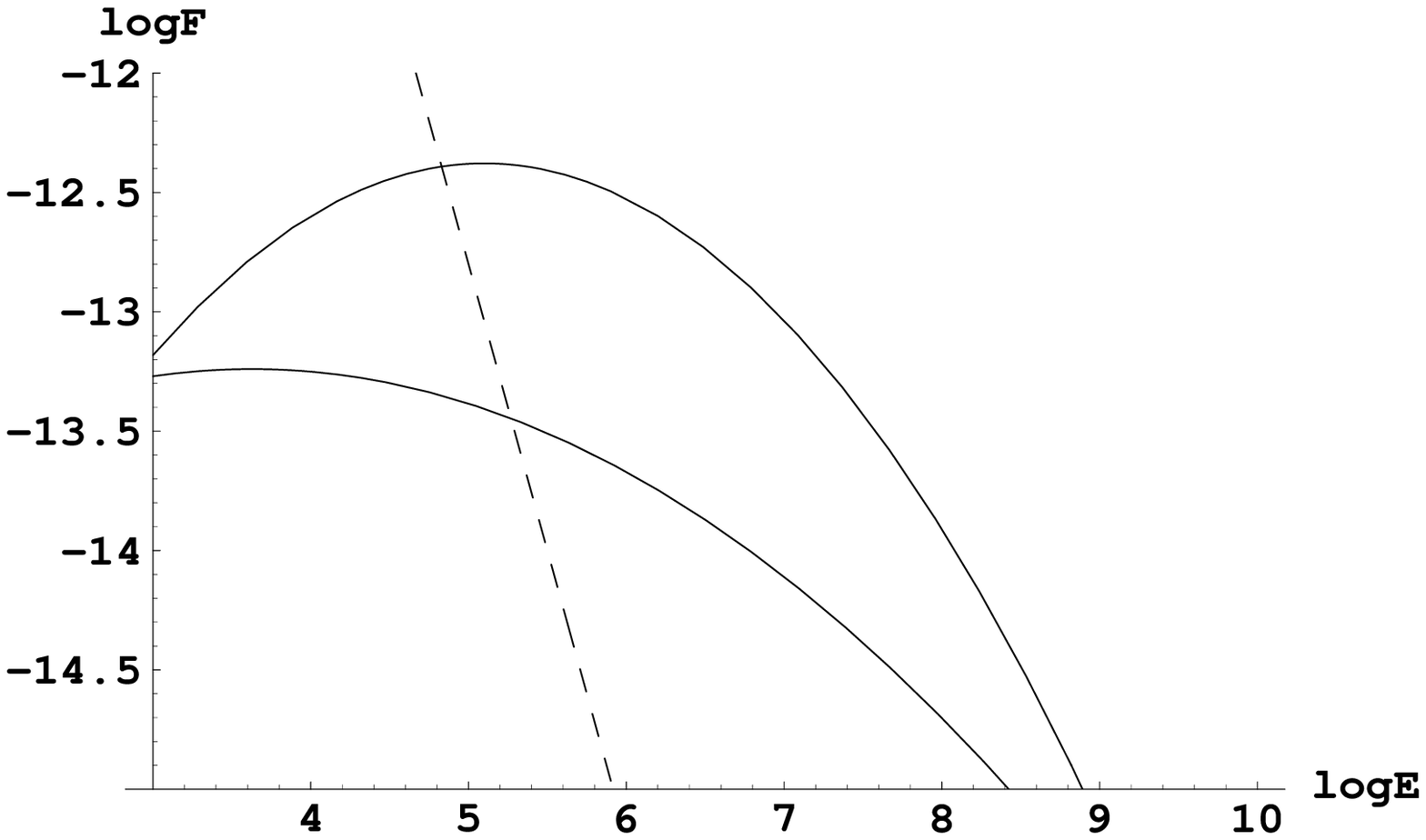}}
\begin{figure}
\centerline{\box4}
\caption{Plot of the expected diffuse $ \nu_\mu + {\bar \nu_\mu}$ flux from
 AGN using the blazar model due to Protheroe (top), the blazar
model due to Mannheim (bottom). The dashed line represents angle averaged
flux from cosmic ray interactions in the atmosphere (see section VI).
Here F is in $ {\rm cm}^{-2} {\rm s}^{-1} {\rm sr}^{-1}$
and neutrino energy E is in GeV.
\label{fig:f.1}}
\end{figure}

\subsubsection{Proton Acceleration, Disk Radiation Interaction}

There is another plausible model \cite{protherb} describing
energetic $\gamma$ ray 
and neutrino emission in blazars which is consistent with gamma ray
observations. In this model protons are accelerated
and directly interact with accretion disk radiation.
This involves only proton acceleration unlike the
previous model which requires the  acceleration of both protons and
electrons.

If the accelerated protons interacted with matter to produce gamma rays
at a time scale consistent with the observed variability of AGN ($\sim
$ 1 day), then the
proton number density must be $ > 10^{9} {\rm cm}^{-3}$. Such a high
density is not normally expected in AGN jets.
Interaction with the radiation field is therefore 
necessary if protons are accelerated, but a sufficiently high 
density of target photons for proton interactions would present
problems for gamma ray escape if the radiation were isotropic. 
To solve this problem 
the $\gamma$ ray emission region is assumed to be
positioned so that the impinging radiation is  highly anisotropic.
The photon spectrum used is a modified version of the standard 
thin accretion disk spectrum \cite{shakura}.

Both blazar models described above find that the $\gamma$ ray spectra 
from hadronic synchrotron cascades can explain the observed properties of 
$\gamma$ ray emitting AGN. 
They also predict very similar neutrino fluxes from the source.
The disk spectrum is expected to be more important if particle
acceleration occurs at distance less than a parsec from the central black
hole and the sychrotron spectrum dominates at larger distances.
If there is continuous proton acceleration along the jet, both target
fields are important.
Figure 1 illustrates the expected diffuse neutrino flux from the two
different blazar AGN models described here in comparison to the
atmospheric neutrinos due to cosmic ray interactions.

\subsubsection{Dominance of Hadronic Acceleration Models}

The hadronic models compete with the leptonic models
in explaining the TeV $\gamma$ ray emission from nearby blazars.
Hadronic emission models naturally predict emission above $\sim$ 10
TeV unlike leptonic models as indicated below.
This can be understood in two ways. Considering first order
Fermi acceleration at shocks in the hadronic blazar model \cite{mannheimb},
the maximum achievable proton and electron energies can be attained
for a given value of the magnetic field by
balancing the proton and electron
energy losses from different processes with energy gains from
Fermi acceleration.
The maximum Lorentz factor for electrons
is found to be,

\begin{equation}
\gamma_{emax} = 4 \times 10^{7} K^{1 \over 2} B^{-{1 \over 2}} f_e^{-{1 \over 2}
}
\end{equation}

\noindent
where B is the associated magnetic field, $K < 0.4$ depends on diffusion
coefficients and $f_e$ typically $\sim$ 2,
depends on different energy loss mechanisms (takes
into account the relative contributions of synchrotron, inverse Compton,
Bethe-Heitler pair production and pion production) \footnote{ $f_e = 1 +
u_\gamma / u_B $ which is typically 2 considering the radiation energy
density $u_\gamma$ to equal the magnetic energy density $u_B $.}.
Now if as an example, the 5-8 TeV emission from Mkn 421 is to be explained
by a purely leptonic model, the electron Lorentz factor must reach at least
$ 10^{7}$. If these same electrons are
assumed to produce the observed X ray spectrum by synchrotron process,
for reasonable parameters a magnetic field of only $10^{-4}$ G is expected
\footnote{A magnetic field of $ 0.1 - 100 $ G is expected in the $\gamma$ 
ray and neutrino emitting zone of the jet \cite{mannheimb}}.
Such a magnetic field is observationally found in the hot spots of 
jets at kiloparsec
scale which is $\sim 10^6$ larger than the size of $\gamma$ ray emitting
zone inferred by observed AGN variablilty. 
This therefore cannot account 
for the $\sim$ 10 TeV emission by purely
leptonic processes.

In a separate model \cite{masta} the maximum
achievable energy of a leptonic photon is also shown to be 10 TeV.
The variable flux of TeV $\gamma$ rays detected from Mkn 421 and Mkn 501
requires the presence of high energy electrons which could in principle
produce a large number of electron positron pairs, leading to an
electromagnetic cascade.
In \cite{masta} it is described how this scenario can be avoided
if electrons are accelerated rectilinearly in an electric field
rather than being isotropically
injected into a blob as in most models of the GeV $\gamma$ ray emission.
If electrons are considered to be subject to rectilinear acceleration
balanced by inverse Compton losses from scattering on accretion
disk photons, the maximum achievable photon energy
is found to be $\sim$ 10 TeV, by the effective exclusion of the possibility
of electromagnetic cascades during acceleration.

A surprisingly strong $\gamma$ ray flux
observed from Mkn 501
extending to 10 TeV \cite{recent} without a well defined cut off
therefore strongly
suggests hadronic models and points to the production of TeV
neutrinos.

\subsection{Spherical Accretion Model for Ultrahigh Energy Neutrino Production}

High energy neutrino
production in an AGN environment can also be described using the
spherical accretion
model~\cite{sph}.
According to this model, close to the black hole the accretion
flow becomes spherical and a shock is formed where the inward pressure of the
accretion flow is balanced by the radiation pressure.
The pions leading to neutrino production are
produced in this model through the collision of
fast protons (accelerated through
first-order diffusive Fermi mechanism at the shock~\cite{sph}) with the
photons of the dense ambient radiation field.
These neutrinos, produced in the central regions of the AGN,
are expected to dominate the neutrino sky
at energies of 1 TeV and beyond.

\subsubsection{UHE Neutrino Emission from the Center of the Milky Way}

Our galaxy harbors a compact non thermal
source named Sagittarius A* (or Sgr A*) whose characteristics suggest
that it may be a massive ($ \sim 10^{6} M_{\odot}$ ) black hole accreting
matter from an ambient wind in that region \cite{center} which is assumed
to originate from IRS 16, a nearby
group of hot massive stars.
X-ray and $\gamma$ ray emissions have been detected from the galactic center.
Recently the EGRET instrument on the Compton Gamma Ray Observatory
 has identified a central continuum source with
luminosity $\sim 10^{36}$ ergs/sec \cite{mattox}.
Considering that a single supermassive black hole could be present
the activity of the galactic center could be similar to the 
activity at the center of an AGN \cite{mastacenter}. 
The shock acceleration mechanism would follow spherical accretion
since there is no observed evidence for a blazar.
Due to the absence of the dense radiation fields, typical of
AGN conditions, the collisions between relativistic and ambient
matter to produce
$\gamma$ rays and neutrinos would be the dominant process. 
All suggested models till now \cite{center} attempt 
to accomodate the observed EGRET data, which
is of relatively low energy ($\sim$ 30 MeV - 10 GeV). 
However the galactic center could be a 
potential source of UHE $\gamma$ ray and neutrino 
emission; future observations
by ground based telescopes e.g. CANGAROO is necessary to
ascertain this.

\section{Neutrinos from Supernova Remnants and Pulsars. }

Supernova remnants (SNRs) could be 
the principal sources of galactic cosmic rays
up to energies of $\sim 10 ^{15}$ eV \cite{snr1}. A fraction of the accelerated
particles interact within the supernova  remnants and produce
$\gamma $ rays.
If the  nuclear component of cosmic rays is strongly enhanced inside
supernova remnants, then $\gamma$ rays and $\nu$s are invariably
produced through nuclear collisions which lead to pion production and
subsequent decay.
Observations of both UHE $\gamma$ ray and $\nu$ from SNR sources would 
suggest accelerated hadrons in SNR.

Recent observations above 100 MeV by the EGRET 
instrument have found $\gamma$
ray signals from several SNRs (e.g. IC 443, $\gamma$ cygni, etc.) \cite{egret}. 
However, the production mechanisms of these high energy gamma-rays 
has not been unambiguously identified. The
emission may be due to the interaction of the SNR blast wave
accelerated protons with adjacent molecular clouds \cite{drury}.
Evidence that this is the dominant process would be the identification of
a pion bump near 70 MeV which has not been seen in the EGRET
SNR data--possibly due to poor statistics. It is also possible
that the high energy $\gamma$ rays arise from pulsars residing 
within the SNRs.

Evidence for electron acceleration in SNRs comes from the ASCA satellite
observation of non-thermal X-ray emission from SN 1006 \cite{koyama}
and IC 443 \cite{keohane}. Ground based telescopes have detected TeV emission
from SN 1006 \cite{cangre} and the Crab Nebula \cite{cangaroo}.
For recent reviews see \cite{snr1}.
Our objective in this paper is to look closely at SNRs as sources
of UHE neutrinos and we investigate different possibilities in the sections
below.

\subsection{Neutrinos from Supernova Remnants Assuming p-p Interactions}

There are two schools of thought describing the acceleration
mechanisms in SNRs. In one, 
TeV $\gamma $ rays are suggested to
be leptonic in origin \cite{leptonic} and it is shown that SNRs 
could produce a TeV photon flux through inverse Compton scattering off the
microwave radiation and other ambient photon fields.
In the second, accelerated protons produce
$\gamma$ rays by the decay of neutral pions produced in 
proton nucleon collisions. We are
interested in the possibilty of TeV neutrino emission from
SNRs which is possible only in hadronic models.
Protons are accelerated by the SNR blast wave, or, if a
pulsar is present within the remnant, acceleration could be
due to pulsar wind terminated shock (termination at the surrounding
nebula).
Non linear particle acceleration concepts
have been used in \cite{drury} considering a general SNR model.
Following \cite{drury} 
the production rate of $\gamma$ rays per unit volume can be written
as,

\begin{equation}
Q_\gamma = E_\gamma n = q_\gamma n E_C
\end{equation}

\noindent
where $n$ is the number density of the gas, $E_C$ is the cosmic
ray energy density, $E_\gamma$ is the $\gamma$ ray emissivity
per unit volume and the production rate $q_\gamma = {E_\gamma \over E_C}$.
Considering  a differential energy spectrum of accelerated
protons inside the remnant of $ E^{2 - \alpha}$,
$q_\gamma$ would be given by,

\begin{equation}
q_\gamma( >E )= q_\gamma(\ge 1 {\rm TeV}) {\left({E \over 1{\rm TeV}} \right)}^{3-\alpha}
\end{equation}

\noindent
Numerical values of $q_\gamma$
for different spectral indices
of the parent cosmic ray distributions are given in Table 1.
The contribution of nuclei other than H in both the target matter
and cosmic rays is assumed to be the same as in the ISM \cite{gamma}.
The units of $q_\gamma$ are ${\rm s}^{-1}\; {\rm erg}^{-1}\; {\rm cm}^{3}\; {\rm(H-atom)}^{-1}$.
The corresponding total $\gamma$ ray luminosity can be estimated by
$\int q_\gamma E_C d^{3}r$ which, for simplified models, can be
written as

\begin{equation}
L_\gamma = \theta q_\gamma E_{\rm sn} n
\end{equation}

\noindent
where $\theta$ is the fraction of the total supernova explosion 
energy (${E}_{\rm sn}$) converted to cosmic ray energy. 
The flux of $\gamma$ rays with energy $>1$ TeV from a SNR at
a distance d from the Earth calculated using $ F_{\gamma} = {L_\gamma
\over {4 \pi {\rm d}^2}}$, can be written as \cite{drury},

\begin{equation}
F_{\gamma}( >{\rm TeV}) \sim 8.4 \times 10^6 \;\; \theta \; q_\gamma (\alpha)
{\left ( {E \over 1 {\rm TeV}} \right )}^{3 - \alpha} \left ( {E_{\rm sn} \over 10^{51}{\rm erg}
} \right ) \left ( {n \over
1{\rm cm}^{-3}} \right ) {\left ( { {\rm d} \over 1{\rm kpc}} \right )}^{-2}
 \;\; {\rm cm}^{-2} {\rm s}^{-1}.
\end{equation}

\noindent

These results correspond to the SNR in the Sedov Phase where the luminosity is
roughly constant. The total luminosity is low during the free expansion
phase and the Sedov phase starts when the amount of swept out matter
equals the ejecta mass. Typically this occurs at a radius $\sim$ 10 pc
and the SNR spends most of its useful life in this phase.
TeV neutrinos can be predicted to be produced
as a byproduct of the 
decay of charged pions.
To find the corresponding UHE neutrino flux ($ F_{{\nu_\mu} + {\bar{\nu_\mu}}}$)
for different
spectral indices  we resort to the calculated ratios 
${F_{{\nu_\mu} + {\bar{\nu_\mu}}} \over F_\gamma} $ \cite{drury,gaisserbook} 
as illustrated in the table below.
$q_\gamma (\alpha)$ values for different  $\alpha$ values
are also included.
A comprehensive discussion of the spectrum weighted moments for secondary
hadrons based on the accelerator beams with fixed targets at beam energies
$\le 1$ TeV has been presented by Gaisser \cite{gaisserbook}. This has also
been shown to characterize correctly the
energy region beyond 1 TeV \cite{drury}. A direct ratio estimate can be 
made as in \cite{drury} to give results
very close to that in \cite{gaisserbook}. For harder spectra the ratio is found
to approach unity.
These values are for particles at the production site. 
The highly penetrating
neutrino flux at the Earth can be estimated to be similar to the one
at the site of production whereas the corresponding photons are readily
absorbed.

\vspace{0.5 cm}
\begin{table}[h]
\begin{center}
\begin{tabular} {|l|c|c|c|c|}
\vspace{0.25 cm}
       $\alpha$ & 4.2 & 4.4 & 4.6 & 4.8 \\
\hline
\vspace{0.25 cm}
${F_{{\nu_\mu} + {\bar{\nu_\mu}}} \over F_\gamma}$ [Gaisser (1990)\cite{gaisserbook}] &0.80&0.67&0.56&0.46 \\ \hline
\vspace{0.25 cm}
${F_{\nu_\mu} + {\bar{\nu_\mu}} \over F_\gamma}$ [Drury {\it et.al.} (1993)\cite{drury}]  &0.86&0.77&0.66&0.58 \\ \hline
\vspace{0.25 cm}
$q_\gamma (\alpha)$ \cite{drury} & $4.9 \times 10^{-18}$ & $8.1 \times 10^{-19}$ &$1.0 \times 10^{-19}$ & $ 3.7 \times 10^{-20}$ \\ 
\end{tabular}
\vspace{0.5 cm}
\caption{Values of expected UHE neutrino and gamma ray ratio at production site.}
\end{center}
\end{table}

We have taken the average of the two ratios as given in
Table 1 to calculate the corresponding
neutrino flux from equation (5) for each spectral index. 
As an example, the expression for the neutrino flux 
 for 
$\alpha \sim 4.2$ would be given by

\begin{equation}
F_{\nu_\mu}( > {\rm TeV}) \sim 3.4 \times 10^{-11} \;\;  \theta {\left ( {E \over 1 {\rm TeV}} \right )}^{-1.2} \left ( {E_{\rm sn} \over 10^{51}{\rm erg}} 
\right ) 
\left ( {n \over 1{\rm cm}^{-3}} \right )
{\left ( { {\rm d} \over 1{\rm kpc}} \right )}^{-2}
 \;\; {\rm cm}^{-2} {\rm s}^{-1}
\end{equation}

\setbox4=\vbox to 160 pt {\epsfysize= 5 truein\epsfbox[0 -200 612 592]{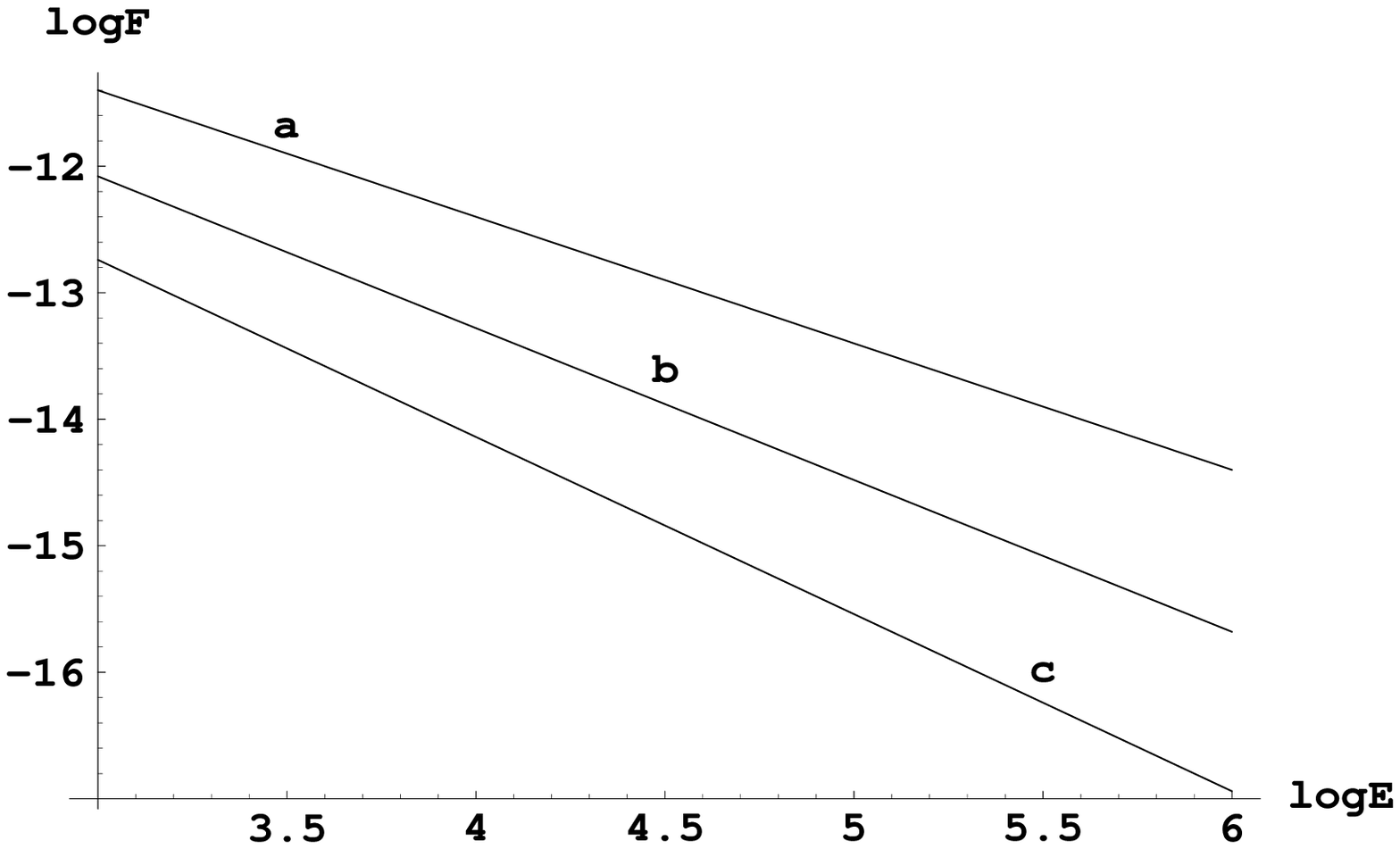}}
\begin{figure}
\centerline{\box4}
\caption{Plot of the expected neutrino flux from the Crab considering
hard source spectra, $\alpha$ $\sim$ 4.0 (a), 4.2(b), 4.4(c).
Here F is in ${\rm cm}^{-2} {\rm s}^{-1}$ and 
neutrino energy E is in GeV.
\label{fig:f.2}}
\end{figure}

Very recent data from the CANGAROO detector \cite{cangaroo} indicate that the
energy spectrum of $\gamma$ rays from the Crab pulsar/nebula (filled center SNR)
extends up to at least 50 TeV. 
The IC (Inverse Compton) $\gamma$ ray spectrum for the Crab has been
calculated by several authors on the basis of the SSC (Synchrotron
Self Compton) model and recently it has been suggested that \cite{aha}
cosmic microwave background and infrared photons emitted from the
dust in the nebula are the main seed photons for the TeV $\gamma$
ray production by the IC process. However, as the energy increases,
IC processes produce a steeper spectrum because of
synchrotron energy loss in magnetic fields.
All models based on electron processes have difficulty
in explaining the observed hard spectrum that extends to
beyond 50 TeV.
This observation may indicate 
that very high energy relativistic protons are accelerated by the pulsar
wind, thereby supporting hadronic models of $\gamma$ ray emission.

The hadronic mechanism for production of $\pi^0$
and hence $\gamma$ rays as described above has been used in \cite{aha}
to explain the UHE photon emission from the Crab.
Nuclear p-p interactions are considered to occur among the
protons accelerated in the nebula.
The spectrum of $\gamma$ rays from $\pi^0$ in this model gives a 
hard differential
spectrum which matches the CANGAROO observations 
thereby lending credence to the hadronic model. 
An approximate expression for the UHE gamma ray spectrum 
from equation (5) could be written  
for spectral index $\alpha$
varying between 4.0-4.5 as \cite{aha},

\begin{equation}
F_{\gamma} ( > {\rm TeV}) \sim 2.0 \times 10^{-3 - {3 \alpha}}\;\; 
\left ( {W_p \over 10^{48}{\rm erg}} \right )
\left ( {n \over 100 {\rm cm}^{-3}} \right )
{\left ( { {\rm d} \over 2 {\rm kpc}} \right )}^{-2}
 {\left ( {E \over 1 {\rm TeV}} 
\right )}^{3 - \alpha} \;\; {\rm cm}^{-2} {\rm s}^{-1}
\end{equation}

\noindent
where d is the distance to the source (distance to Crab is 2 kpc),
$W_p$ is the kinetic energy of relativistic accelerated protons (reasonable
value for the Crab would be $\sim 10^{48}$ erg) 
and $ n$ is the effective number density
($\sim$ 100 ${\rm cm}^{-3}$). 
The corresponding UHE neutrino flux can be calculated directly
using values from Table 1. For reasonable parameters 
for $\alpha \sim 4.2$ as an example the neutrino flux from the Crab would be, 

\begin{equation}
F_{\nu} ( > {\rm TeV}) \sim 8.3 \times 10^{-13}
{\left ( {E \over 1 {\rm TeV}} \right )}^{-1.2} \;\; {\rm cm}^{-2} {\rm s}^{-1}
\end{equation}

\noindent
Figure 2 illustrates the  expected neutrino flux from 
the Crab Nebula for different spectral indices as calculated above.
The spectra are normalized such that
$ A =  \left ( {W_p \over 10^{48}{\rm erg}} \right )
\left ( {n \over 100 {\rm cm}^{-3}} \right ) 
{\left ( { {\rm d} \over 2{\rm kpc}}\right )}^{-2} \; = 1 $.
SNR can therefore be considered to be effective sources for UHE neutrino
production. More details can be found in \cite{mou1}.

\subsection{UHE neutrinos from SNR and pulsars due to Nuclear Interactions}

It is widely believed that the nuclear component of cosmic rays is
produced in supernova remnants, at least for particle energies
less than about $10^{14}$ eV per nucleon. 
However, there is no direct observational evidence for accelerated nuclei
in SNR. 
In an effort to study the importance of nuclear interactions to the
possible production of UHE neutrinos  from astrophysical 
point sources, we turn to a model proposed by Protheroe
{\it et. al.} \cite{protheroesnr}. 
In this model very young SNR are considered in which 
nuclei, mainly Fe, extracted from the neutron
star surface and accelerated to high Lorentz factors are
photodisintegrated during propagation through the
neutron star's
radiation field. Photodisintegration can also occur in the 
presence of extremely
strong magnetic fields typical of neutron star environments ($\sim 10^{12}$ G).
For acceleration to sufficiently high energies we need a short
initial pulsar period ($\sim$ 5 ms).
The energetic neutrons produced as a result of photodisintegration 
interact with target
nuclei in the shell as they travel out of the SNR, 
producing gamma ray
and neutrino signals; those neutrons passing through the shell 
decay into relativistic
protons, contributing to the pool of galactic cosmic rays.
For a beaming solid angle to the Earth of $\Omega_b$, 
the neutrino flux in this model can be calculated from,

\begin{equation}
F_\nu(E_\nu) \sim  { {\dot N}_{\rm Fe} \over {\Omega_b d^2}} [ 1 - {\rm exp}(
-\tau_{pp})] \int N_n (E_n) P_{n\nu}^M (E_{\nu},E_n) d E_n
\end{equation}

\noindent
where ${\dot N}_{\rm Fe}$ is the total rate of Fe nuclei injected,
$d$ is the distance to the SNR, $P_{n\nu}^M (E_{\nu},E_n) d E_n$ is the
number of neutrinos produced with energies in the range $E_{\nu}$ to
$(E_{\nu}+ d E_{\nu})$ (via pion production and subsequent decay),
and $ N_n (E_n)$ is the spectrum of neutrons extracted
from a single Fe nucleus. $\tau_{pp}$ is the optical depth of the
shell to nuclear collisions
(assuming shell type SNR) which is a function of the mass ejected into the
shell during the supernova explosion and of the time after explosion.
Figure 3 (solid line) gives the $\nu_\mu + {\bar \nu_\mu}$ spectra
obtained from this model at a distance of 10 kpc with  the time
after explosion being 0.1 year.
Signals from nuclei not completely fragmented are ignored because
these particles are charged and would be trapped in the central
region of the SNR which has a relatively low matter density 
and therefore would not make any significant contribution to
neutrino fluxes.

\setbox4=\vbox to 160 pt {\epsfysize= 5 truein\epsfbox[0 -200 612 592]{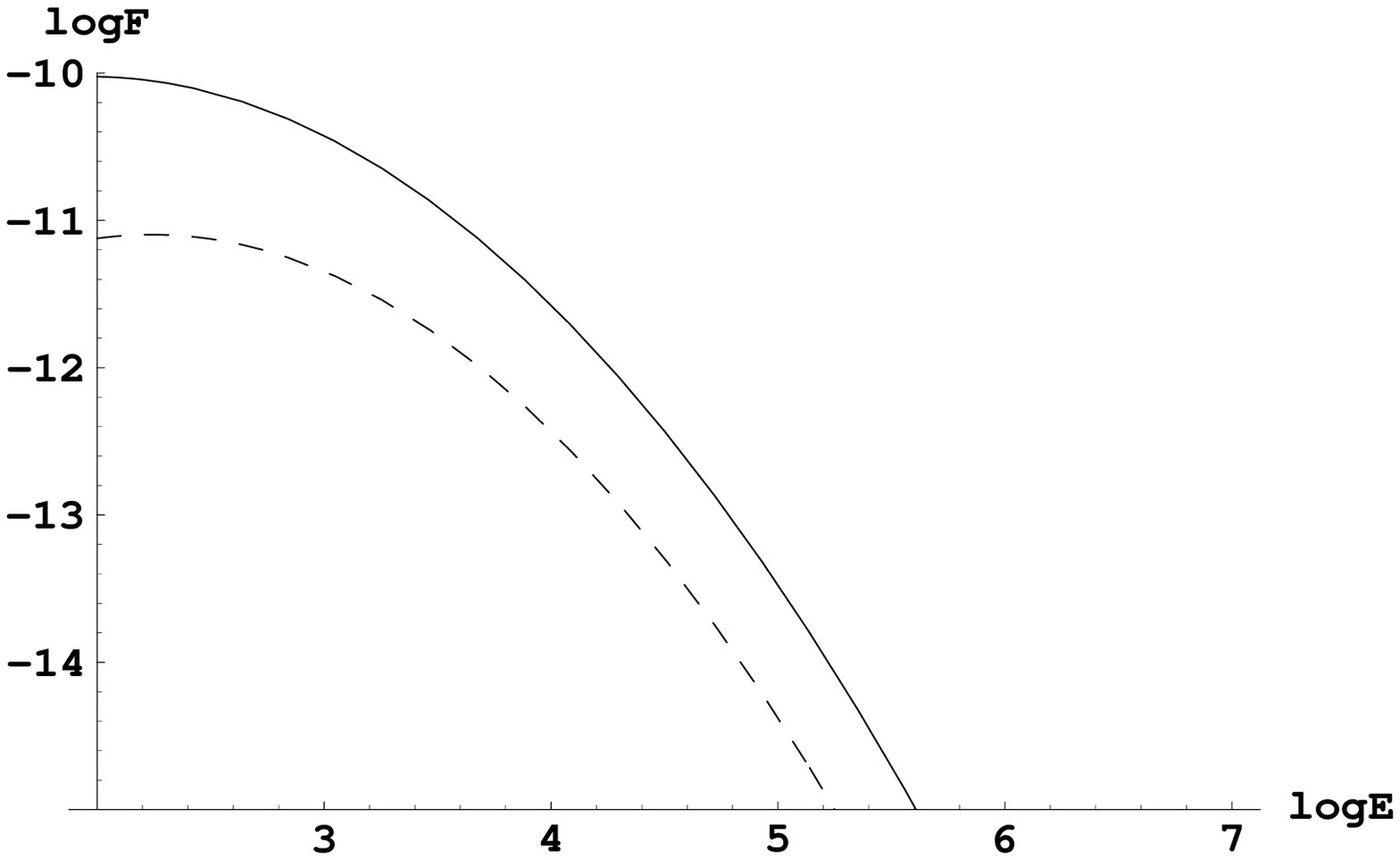}}
\begin{figure}
\centerline{\box4}
\caption{Plot of the expected neutrino $ \nu_\mu + {\bar \nu_\mu} $
spectrum from SNR produced by collisions of neutrons with matter in a
supernova shell for B = $10^{12}$ G and initial pulsar period
5 ms using the maximim polar cap heating
model (very young SNR) (solid line), specifically for Crab Nebula 
using a similar model (dashed line);
Here F is in ${\rm cm}^{-2} \;\; {\rm s}^{-1}$  and neutrino energy E is in GeV.
\label{fig:f.3}}
\end{figure}

Energetic radiation from the Crab Nebula could also be a consequence
of acceleration of heavy nuclei in the pulsar magnetosphere
\cite{protheroepuls}.
Accelerated nuclei can photodisintegrate in collisions with soft
photons produced in the pulsar's outer gap, injecting energetic neutrons which
decay either inside or outside the Crab Nebula. 
The protons from neutron decay inside the nebula are trapped 
by the Crab magnetic field ($\sim 10^{12}$ G)
and accumulate inside the nebula producing gamma rays and neutrinos 
in collisions with matter in the nebula.
The nuclei which survive are confined to the inner parts of the
Crab Nebula which has no evidence for dense matter and
hence neutrino production is negligible.
Figure (3) (dashed line) shows the predicted spectrum from
Crab Nebula assuming this model for some reasonable
pulsar parameters.
 
\section{Neutrinos Associated with Gamma Ray Bursts: Cosmological Scenarios}

Gamma ray bursts (GRB) are presently the most enigmatic astrophysical
phenomenon. Recent observations indicate that they originate from
cosmological sources.
Here we look at the only two GRB models which predict
neutrinos of energy $\ge$ 1 TeV, namely the 
ultra-relativistic fireball model \cite{bahcall}
and the cosmic string model \cite{plaga}. 
Neutrinos from GRBs could be used to test the simultaneity
of neutrino and photon arrival to an accuracy of $\sim$ 1ms,
checking the assumption of special relativity that photons and
neutrinos have the same limiting speed. These observations would also
test the weak equivalence principle according to which photons and
neutrinos should suffer the same time delay as they pass through
a gravitational potential.

\subsection{Ultra-relativistic Fireball Model }

General phenomenological considerations indicate that GRBs could be
produced by the dissipation of the kinetic energy of a relativistic expanding 
fireball \cite{piran} \footnote{ The overall idea of a fireball
model is that a large amount of energy is released in a compact
region of radius $ R \sim c \Delta t$ which is opaque to photons.
The system turns into a fireball
with dense radiation and $e^+$ $e^-$ pair fluid. The fluid then expands
and cools adiabatically in the process and photons escape only
when the optical depth of the fireball is sufficiently reduced.}.
Until recently neutrinos expected as byproducts of GRBs from the fireball model
have been predicted to have energies of $\sim 100$ MeV \cite{halzen1}.
According to Waxman and Bahcall \cite{bahcall} however,
a natural consequence of the dissipative cosmological 
fireball model of gamma ray
bursters is the conversion of a significant fraction of fireball energy
into an accompanying burst of $ \sim 10^{14} $
eV neutrinos, created by photomeson production of pions in interactions
between the fireball $\gamma$ rays and accelerated protons.

The basic picture is that of a compact source producing a relativistic wind.
The variability of the source output results in fluctuations of the
wind bulk Lorentz factor which leads to internal shocks in the
ejecta. Both protons and electrons are accelerated at the shock
and $\gamma$ rays are radiated by synchrotron and inverse Compton
radiation of shock accelerated electrons.
The observed GRB photon spectrum 
is well fitted by a broken power law, $ d N_{\gamma} / d E_{\gamma}\; \sim \;
E^{-{\beta}}_{\gamma} $ with the break energy $E_{\gamma b}$
(where $\beta$ changes from 1 to 2) typically being $\sim 1$ MeV in the range 30 KeV - 3 MeV (BATSE range).
The interaction of protons accelerated to a power law distribution
$ d N_p / d E_p\; \sim \;E^{-2}_p $ with GRB photons results in a
broken power law neutrino spectrum similar to the gamma ray spectrum
\cite{bahcall}.
The accelerated protons undergo photomeson interactions and produce
a burst of neutrinos to accompany the GRB.
The neutrino break energy $E_{\nu b}$ is fixed by the
threshold energy of protons for photo-production in interaction with
dominant $\sim 1$ MeV photons in GRB \footnote {The fractional energy loss
rate of a proton due to pion production is calculated taking into account
$\Delta$ resonance effects. The corresponding fraction of energy
lost in the expanding fireball is found to be maximum when the 
proton energy is $\sim 10^{16}$ eV
dominantly interacting with a 1 MeV photon.}.
If GRBs are sources of ultra
high energy cosmic rays, assuming the sources
are cosmologically distributed the expected neutrino flux from this model 
\cite{bahcall} would be

\begin{equation}
F_\nu \sim 1.5 \times 10^{-9} \; \left ( {{f_\pi} \over 0.2} \right ) 
\; \left ( {1 {\rm GeV} \over E_{\nu}} \right ) \; {\rm min} \left (1,
E_{\nu}/E_{\nu b} \right )
{\rm cm}^{-2} \;\; {\rm s}^{-1} \; \; {\rm sr}^{-1}
\label{aaaa}
\end{equation}

\noindent
where $E_{\nu}$ is the neutrino energy and the neutrino break energy
$E_{\nu b}$ is given by,

\begin{equation}
E_{\nu b} \sim 5 \times 10^{14} \Gamma^2_{300} (E_{\gamma}^b/1{\rm MeV})^{-1} {\rm eV}, 
\end{equation}

\noindent
where $ \Gamma_{300} = {\Gamma \over 300} $, $\Gamma$ being the wind bulk 
Lorentz factor typically 300.
The normalization of the flux depends on the efficiency of pion
production. In equation (\ref{aaaa}), $f_\pi$ is the fraction of 
energy lost to pion 
production by protons
producing the neutrino flux 
above the break, which is essentially independent of
energy and is given by,

\begin{equation}
f_{\pi} = 0.20 \; {L_{\gamma 51} \over (E_{\gamma}^ b/ 1{\rm MeV})\; 
\Gamma^4_{300}\; \Delta t}
\label{aaa1}
\end{equation}

\noindent
where $ L_{\gamma}$, the average 
source luminosity $ = { L_{\gamma} \over 10^{51} }$, and 
$\Delta t$ the time scale of the variablity of the
source is typically $10^{-3}$ s.
For typical GRB producing parameters therefore $f_{\pi} = 0.2$.
From (\ref {aaaa}), the neutrino flux at the neutrino break energy and below
is found to be,
$ \sim 1.5 \times 10^{-14}\; {\rm cm}^{-2} \;\; {\rm s}^{-1} \; \; {\rm sr}^{-1}$.
At still higher energies ($>10^{16}$ eV), the synchrotron losses are the
dominant effect suppressing the neutrino flux \cite{bahcall}.
For interaction with a 1 MeV photon the pion production threshold proton
energy is $\sim 10^{11}$ eV \cite{stecker}
and so the expected neutrino flux drops off
at lower energies.
In this model therefore, most of the neutrino energy is carried by neutrinos
with energy close to the break energy, $\sim 10^{14}$ eV. 

\setbox4=\vbox to 160 pt {\epsfysize= 5 truein\epsfbox[0 -200 612 592]{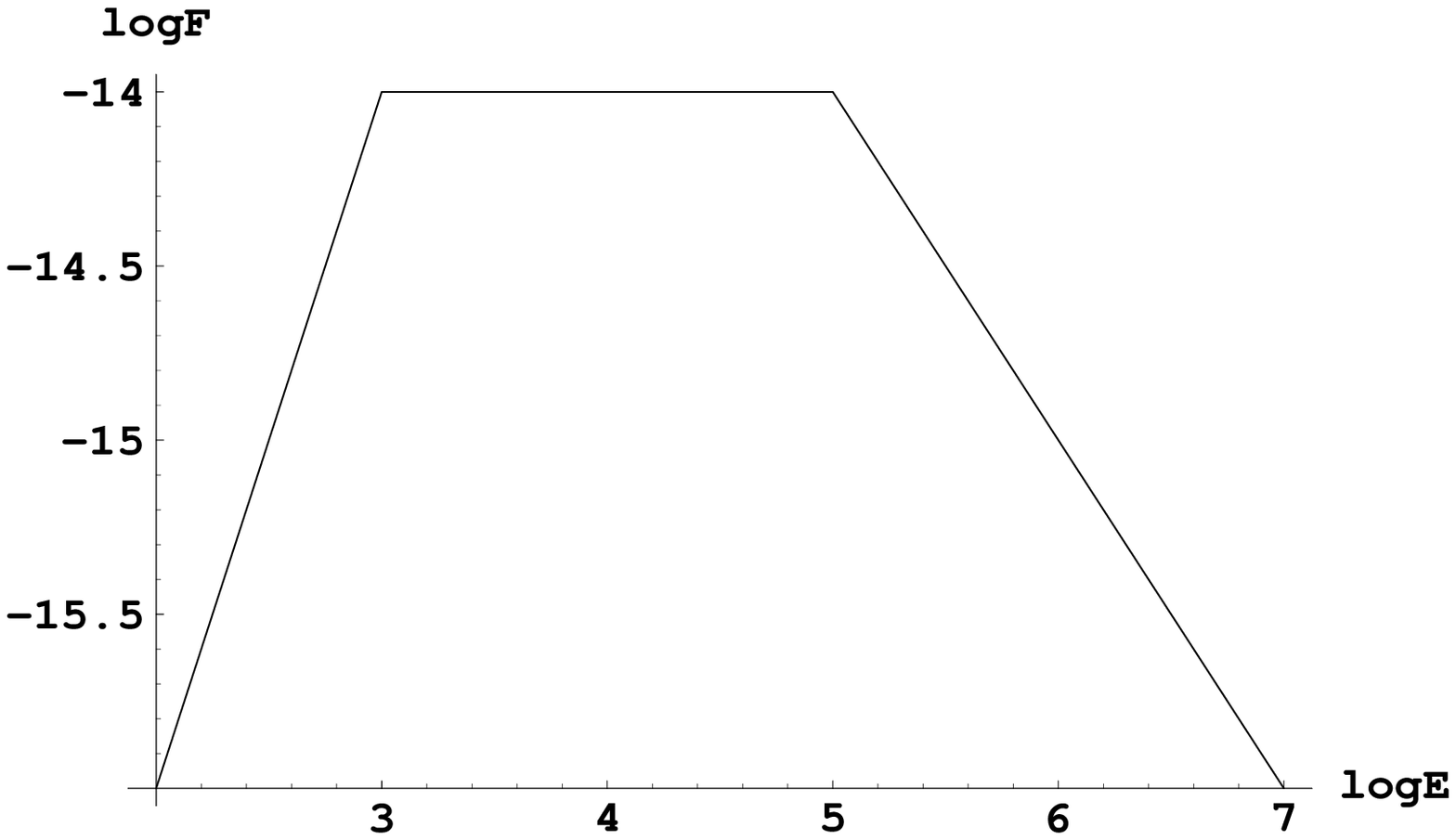}}
\begin{figure}
\centerline{\box4}
\caption{Plot of the expected $ \nu_\mu + {\bar \nu_\mu}$ flux from the
Relativistic Fireball Model.
Here F is in ${\rm cm}^{-2} \;\; {\rm s}^{-1} \; {\rm sr}^{-1}$ and neutrino
energy  E has units of GeV.
\label{fig:f.4}}
\end{figure}

\noindent
Figure 4 illustrates the expected neutrino flux from
the GRB model described above.
These neutrino bursts should be easily detected above the 
background due to atmospheric neutrinos, since they would be correlated both in 
time and angle to the GRB $\gamma$ rays. 

\subsection{ Cosmic String Model of neutrino production}

Cosmic strings are topological relics from the early universe which
could be superconducting and carry electric current under certain circumstances.
A free string (a nonconducting string uncoupled from electromagnetic
and gravitational fields) generically attains the velocity of light
at isolated points in time and space, which are known as cusps \cite{strings1}.
Superconducting cosmic
strings (SCS) emit
energy in the form of classical electromagnetic radiation and ultra-heavy
fermions or bosons which decay or cascade at or near the cusp.
Using recent progress on the nature of electromagnetic symmetry
restoration in strong magnetic fields, the study of the decay 
products of ultraheavy
fermions near SCS cusps consistent with an SCS explanation of $\gamma$
ray bursts shows that the energy emitted from the cusps is found to be mostly
in the form of high energy neutrinos
\cite{plaga,strings1}. The neutrino flux
is roughly nine orders of magnitude higher than that of the $\gamma$ rays.
The flux for a typical neutrino burst using this model is given
approximately by \cite{halzen1},

\begin{equation}
N_\nu = 10^4 \left ({ 10^{-10} \over {\eta_\gamma}}\right ) \left ( {{\rm TeV} 
\over {E_\nu}} \right ) 
\left ({F_{\gamma} \over {10^{-9} {\rm J}\;\; {\rm m}^{-2}}} \right )
{\rm cm}^{-2} \;\; {\rm s}^{-1}
\end{equation}

\noindent
where ${\eta_\gamma}$ is the fraction of energy lost to $\gamma$
rays which is very small with values $ \sim 10^{-9}-10^{-10}$.
$E_\nu$ is the corresponding neutrino 
energy in TeV, and $ F_{\gamma} $ is the observed $\gamma$ ray
flux for a typical burst which can be taken to be $ \sim 10^{-9} {\rm J}\;\; 
{\rm m}^{-2} $.
A simple calculation yields,
$N_\nu \sim 10^{4} \times {E_\nu}^{-1} \;\; {\rm cm}^{-2} \;\; {\rm s}^{-1}$.
If the observed neutrinos are of energy
$\sim 10^{14}$ eV as predicted by the previous model, it can be seen
that the flux for a typical neutrino burst considering the cosmic string
model is much larger being $ \sim 10^{2} \; {\rm cm}^{-2} \;\; {\rm s}^{-1}$. 
Neutrinos predicted by this class of models
have a large energy range, 
from $\sim$ MeV to TeV, in contrast to the fireball model
which predicts abundant neutrino production only up to the
break energy.

This is therefore another model that predicts high energy neutrinos to be
observed in coincidence with $\gamma$ ray bursts. 
If this model is true, TeV neutrinos from such sources are most likely
to be seen and cannot escape the scrutiny of detectors.

\section{Neutrinos from Topological Defect Models}

Topological defect models can directly predict UHE neutrino production.
The main problem with these models is the wide range of model
parameters in which these scenarios could be applied. In all other
models described above the parameters are restricted by observation.
Due to the uncertainty there appears no direct 
reason to invoke these models, though
the theoretical predictions cannot be ignored.

In the conventional topological defect model, the network of long strings
(curvature radius greater than the Hubble radius)
lose their energy predominantly by  gravitational radiation
resulting in very small high energy particle flux. As an extension to
this \cite{topo1} describes
a model where particle production is the dominant energy loss mechanism
in which case therefore, the predicted flux of high energy particles 
would be much
larger.

Physics beyond the standard model might imply the cosmological production
of particles with grand unification scale energies.
In \cite{topo2} a class of models has been explored of exotic heavy
particle decay that ultimately leads to UHE neutrinos. In particular,
UHE neutrinos are produced by the direct decay of a supermassive
elementary ``X" particle associated with some grand unified theory (GUT).
These particles may arise from the collapse of
networks of ordinary cosmic strings or from annihilation of magnetic
monopoles. The predicted flux is very model independent and found
to be smaller in comparison with the model mentioned above.

\setbox4=\vbox to 160 pt {\epsfysize= 5 truein\epsfbox[0 -200 612 592]{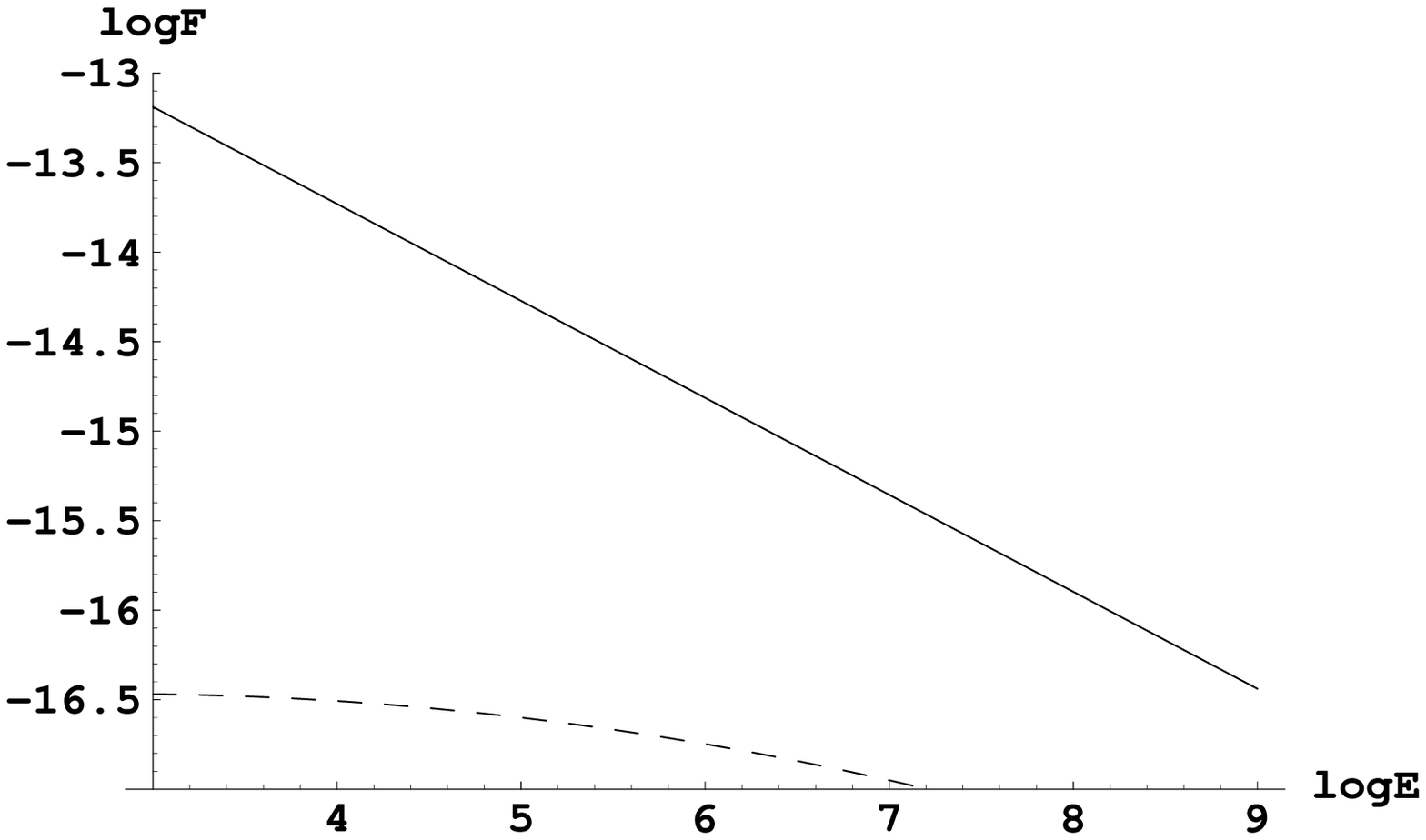}}
\begin{figure}
\centerline{\box4}
\caption{Plot of the expected $ \nu_\mu + {\bar \nu_\mu}$ flux from the
Modified Topological Defect Model (solid) and due to the Exotic Particle
Decay Model (dashed).
Here F is in ${\rm cm}^{-2} \;\; {\rm s}^{-1} \; {\rm sr}^{-1}$ and neutrino
energy  E has units of GeV.
\label{fig:f.5}}
\end{figure}

Figure 5 gives a plot of the expected neutrino flux from the modified 
conventional
topological defect model \cite{topo1}
and also the expected neutrino flux from \cite{topo2}
due to the decay of the exotic X particle.

\section{Overview and Conclusions}
   
A neutrino can be described as a standard Dirac or Majorana particle.
It is well established that neutrino interactions in nature take
place precisely according to the standard electroweak model.
However, whether or not neutrinos have mass is one of the key
questions of particle physics today. 
Neutrinos can be used to
probe the core of some of the most interesting cosmological objects. Due
to their small interaction 
cross sections these particles can stream out unaffected
from even the most violent environments such as those present in Active
Galactic Nuclei (AGN). The presence of several neutrino flavors and spin
states could modify this picture: in their trek from their source to the
detector the neutrinos can undergo flavor and/or spin oscillations which
could obscure some of the features of the source. Because of this, and due
to the recent interest in neutrino astronomy 
\cite{detector}, it becomes important to
understand the manner in which these flavor-spin transitions can occur,
in the hope of disentangling these effects from the ones produced by the
properties of the source. Without such an understanding it will be
impossible to determine the properties of the UHE neutrino source
 using solely the
neutrino flux received on Earth. Incorporation of all oscillation
effects for different UHE neutrino sources and corresponding expected neutrino
fluxes can be found in \cite{hank}.

The study of UHE neutrinos would be incomplete without the mention of
atmospheric neutrino background. Neutrinos produced by cosmic ray
interactions in the atmosphere dominate other sources for neutrino
energy below a few TeV. This is why we focus on neutrinos with
higher energy. The
``conventional" atmospheric neutrino flux \cite{conventional}
is derived from the decay of
charged pions and kaons produced by cosmic ray interactions in the atmosphere.
The atmospheric neutrino flux is  large at 1 TeV but falls with energy
as $E^{-4}$.
The angle averaged atmospheric flux can be parametrized 
\cite{raj} by the equation

\begin{equation}
F_\nu = 7.8 \times 10^{-8}{\left( {E_\nu \over 1 {\rm TeV}} \right)}^{-2.6}
{\rm cm}^{-2} \;\; {\rm s}^{-1} \; {\rm sr}^{-1}
\end{equation}

An order of magnitude calculation indicates that at 1 TeV the flux is
$\sim 10^{-8}{\rm cm}^{-2} \;\; {\rm s}^{-1} \; {\rm sr}^{-1}$, at 10 TeV
it is $\sim 10^{-10}{\rm cm}^{-2} \;\; {\rm s}^{-1} \; {\rm sr}^{-1}$ and
this goes down to $\sim 10^{-13}{\rm cm}^{-2} \;\; {\rm s}^{-1} \; {\rm sr}^{-1}$ at 100 TeV. For a few TeV energies therefore neutrinos from other sources
as discussed in this paper will dominate.
If we raise the energy threshold to 100 TeV the atmospheric neutrino
background is essentially eliminated.
An additional ``prompt" contribution of neutrinos \cite{prompt}
to the atmospheric flux arises from charm production and decay.
The vertical prompt neutrino flux has been recently reexamined using the
Lund model for particle production and has been shown to be small
compared to the conventional flux for energies $\le 10^5$ GeV.
Since atmospheric neutrinos are important only for energies $\le$ a few TeV
neutrinos from charm decay can be neglected in our event rate calculations.
Also the vertical muon flux
could be an unavoidable background. By deploying a detector at
great depths or observing upward going muons or both one can reduce the
cosmic ray muons to a manageable background.

UHE neutrinos can be detected by observing muons, electrons
and tau leptons produced
in charged-current neutrino nucleon interactions \cite{raj,pakvasa,hill}.
The primary means of detection of muon neutrinos and antineutrinos
is by charged current conversion into muons and antimuons.
To minimize the effects of atmospheric muon and neutrino background,
one observes upward going muons.
To observe $\nu_e$ one looks at the contained event rates for
resonant formation of
$ W^-$ in the $ \bar{\nu_e}$ interactions at $E_{\nu} = 6.3$ PeV
for downward moving $\nu_e$.
Also the rapid development of electromagnetic showers may make it possible
to detect upward going air showers initiated by an electron neutrino that
interacts near the earth surface.
In all the models discussed above there is negligible Tau neutrino
flux, hence the observation of Tau neutrinos by the neutrino detectors
would indicate oscillations of the neutrinos, thereby confirming
physics beyond the standard model.
The key signature for the detection of $\nu_\tau$ is the charged current
$\nu_\tau$ interaction, which produces a double cascade on either end 
of a minimum ionizing track \cite{pakvasa}.
The threshold energy for detecting these neutrinos however is near 1 PeV
at which these cascades are separated by roughly 100m which is easily
resolvable in the planned neutrino telescopes.

We have summarized the different models of astrophysical point sources for 
TeV neutrinos in this paper.
A complete and elaborate calculation of expected event rates from UHE
muon and electron neutrinos
looking at selected
AGN models has been carried out
using new neutrino cross section values (taking into account most recent
parton distributions) in \cite{raj} for a detector area of 0.1 ${\rm km^2}$.
However a more detailed study of the expected neutrino detection rates by the
proposed ice/water detectors 
should incorporate all the astrophysical sources  described above
and also include the possibility of detection of Tau neutrinos (adding the
effects of neutrino oscillations). This work is in preparation \cite{hank}.

To show that neutrinos from the above sources are detectable we can make
an estimate looking solely at muon neutrino detection.
In that case the expected event rate for a detector
with effective area A is given by \cite{raj},

\begin{equation}
{\rm Rate} = {\rm A} \int dE_\nu P_\mu (E_\nu; E_\mu^{\rm min}) S(E_\nu) {dN
\over dE_\nu}
\end{equation}

\noindent
where $P_\mu (E_\nu; E_\mu^{\rm min})$ is the probability that a neutrino
of energy $E_\nu$ produces an observable muon with threshold energy
$E_\mu^{\rm min}$, ${dN \over dE_\nu}$ is the diffuse neutrino flux
and $S(E_\nu)$ is the shadow factor of the earth.
Order of magnitude calculations of the astrophysical sources discussed
above show the fluxes to be detectable by kilometer scale detectors.
Let us restrict ourselves as an example to ${\rm km}^2$ detector area,
neutrino energy 100 TeV and muon threshold energy 10 TeV. 
Considering AGN neutrinos the expected event rates would be approximately
6 events/sr/yr and  38 events/sr/yr for the two blazar models
respectively whereas the optimistic spherical accretion model would expect
$10^4$ events/sr/yr (Section II). The SNR models would predict $\sim$ 3
events per year from the Crab considering hard source spectrum
which necessitates a detector with high angular resolution (Section III). 
The GRB model due to Bahcall would predict about 2 events/sr/yr;
though the flux is low this prediction
is advantageous because observed neutrinos can be 
correlated in time and position
with GRBs (Section IV). The Cosmic String Model of GRB which is the most
optimistic of all these models would expect
$\sim 10^{4}$ events/sec (Section IV) and the 
 topological defect models predict about 1 event per year
(Section V). The last two models are dependent on the choice of
parameters not restricted by observations and are therefore uncertain.

The field of high energy neutrino astronomy is looking forward
to the 
construction and development of the 
next generation km$^3$ scale neutrino detector
in water or in ice. Several initiatives are underway namely 
AMANDA, Baikal, NESTOR,
ANTARES and the water km$^3$ scale initiative at
Berkeley. For neutrino astronomy to be a viable science several of 
these, or other projects will have to succeed. Astronomy, whether on the
optical or in any wave band thrives on the diversity of complementary 
instruments, not on a single best instrument. The important fact remains
in the effort to uncover the vast amount of Physics behind the
successful operation of any or all of the efforts to detect neutrinos
in such high energy ranges.

\section{Acknowledgements}

M.R. wishes to thank Dipen Bhattacharya for valuable astrophysics
input. This research has been supported in part by Grant NAG5-5146.

\end{document}